\documentclass[12pt,english]{article}
\usepackage[T1]{fontenc}
\usepackage[latin1]{inputenc}
\usepackage{a4wide}
\usepackage{graphicx}
\usepackage{setspace}
\makeatletter

\usepackage{babel}
\makeatother
\begin{document}
{\huge \noindent Positive Feedback, Stochasticity and Genetic Competence}{\huge \par}

{\Large \noindent Rajesh Karmakar and Indrani Bose$^{*}$}{\Large \par}

\noindent Department of Physics 

\noindent Bose Institute

\noindent 93/1, Acharya Prafulla Chandra Road, Kolkata-700 009, India.

\noindent $*$ Author to be contacted; e-mail: indrani@bosemain.boseinst.ac.in

\begin{abstract}
\noindent A single gene, regulating its own expression via a positive
feedback loop, constitutes a common motif in gene regulatory networks
and signalling cascades. Recent experiments on the development of
competence in the bacterial population \emph{B. subtilis} show that
the autoregulatory genetic module by itself can give rise to two types
of cellular states. The states correspond to the low and high expression
states of the master regulator ComK. The high expression state is
attained when the ComK protein level exceeds a threshold value leading
to a full activation of the autostimulatory loop. Stochasticity in
gene expression drives the transitions between the two stable states.
In this paper, we explain the appearance of bimodal protein distributions
in \emph{B. subtilis} cell population in the framework of three possible
scenarios. In two of the cases, bistability provides the basis for
binary gene expression. In the third case, the system is monostable
in a deterministic description and stochasticity in gene expression
is solely responsible for the appearance of the two expression states.
\end{abstract}

\section{Introduction}

Positive feedback loops are common motifs in gene transcription regulatory
networks and signaling cascades. The simplest such motif is the autoregulatory
loop in which the proteins synthesized by a gene stimulate the production
of more proteins in an autocatalytic fashion \cite{key-1,key-2,key-3}.
In most cases, the concerned gene is also expressed at a basal level,
i.e., proteins are synthesized even when the positive feedback is
non-functional. The autoregulatory dynamics have a nonlinear character
and this combined with positive feedback may give rise to binary gene
expression in a range of parameter values. The protein levels, as
a result, have a bimodal distribution in a population of cells. In
a fraction of cells, the protein level is low and in the rest of the
population the level is high. Theoretical predictions of binary gene
expression have been verified in experiments on single gene autocatalytic
modules in bacteriophage $\lambda$ and \emph{S. cerevisiae} \cite{key-2,key-4}\emph{.}
The experimental findings further suggest that the observed bimodality
has a stochastic origin. A remarkable example of population heterogeneity,
brought about by a combination of autoregulatory positive feedback
and stochasticity, is provided by the bacterial population \emph{B.
subtilis} in which a fraction of the population develops genetic competence.
Microorganisms like bacteria have to cope with a multitude of antagonistic
agents and environmental conditions in order to live. Under such circumstances,
the bacteria may adopt a number of strategies to optimize their chances
of survival \cite{key-1,key-5}. One such strategy is the development
of genetic competence, observed in some bacterial organisms. In the
competence state, specialized proteins are synthesized which allow
the cell to take up large pieces of DNA from the environment and incorporate
them into the bacterial genome. New traits are thus acquired from
genetically distinct organisms. Experiments show that only a small
fraction of the bacterial population reaches the competence state.
The resulting phenotypic diversity in the population may prove to
be advantageous. The individual cells in a homogeneous population
share the same fate when subjected to harmful influences. Diversity
enhances the chance that a fraction of the population, even if small,
is able to survive and adapt to the changed circumstances. In \emph{B.
subtilis}, the development of competence is regulated by the transcription
factor ComK synthesized by the comK gene. The protein functions as
a master regulator which activates the transcription of several genes
including those necessary for DNA uptake. The ComK activity in turn
is controlled by a host of other proteins. An autoregulatory positive
feedback module forms the core of the complex regulatory network.
ComK binds to the promoter of its gene and promotes its own production.
The positive feedback gives rise to bimodality in the cell population
with low and high comK expression states as the stable states. In
the competence state, the level of ComK proteins is high enabling
ComK to act as a transcription factor. Two independent experiments
\cite{key-6,key-7} have confirmed that an autostimulatory loop of
comK expression is by itself sufficient to establish competence bimodality
in a bacterial culture. The experimental findings moreover suggest
that stochasticity plays an essential role in the establishment of
competence. 

In this paper, we study a simple model of autoregulatory positive
feedback involving a single gene using both deterministic and stochastic
descriptions. In the deterministic case, positive feedback and nonlinearity
result in bistability in a range of parameter values. The two stable
steady states correspond to low and high gene expression levels. The
high expression state is reached when the protein level exceeds a
threshold value. The autocatalytic switch is then triggered bringing
about a full activation of the autostimulatory loop. In the absence
of such activation the proteins are synthesized at a low level. Bimodality
in a cell population requires the autocatalytic switch to be triggered
in a fraction of the cell population. This is where stochasticity
in gene expression comes into the picture. Several recent studies,
both theoretical and experimental, highlight the significant role
of stochasticity in gene expression and its regulation \cite{key-8,key-9,key-10,key-11,key-12}.
The two stable steady states are separated by an unstable steady state.
The corresponding protein level (intermediate level of gene expression)
provides the threshold for the triggering of the autocatalytic switch.
The low (high) expression state is obtained when the protein level
is below (above) the threshold value. Stochasticity in gene expression
gives rise to fluctuations in the protein levels and the fluctuations,
if sufficiently large, bring about transitions across the threshold.
In the deterministic picture, bifurcations occur at two special values
of the parameter $J_{0}$, the rate for basal protein synthesis. At
the lower (upper) bifurcation point, there is a transition from monostability
(bistability) to bistability (monostability). This framework provides
an alternative explanation of population heterogenity. Inducer molecules
are often required to initiate gene expression at the basal level.
The distribution of the molecules may be non-uniform in a population
of cells. Thus, the basal levels in the individual cells are not identical
but have a disribution around an average value. If this distribution
overlaps with the upper bifurcation point, the cell population develops
a bimodal character. There is also a third explanation for population
heterogeneity which is solely based on stochasticity in gene expression.
In this case, the system is not bistable in the deterministic picture
and bimodality occurs due to random transitions between the low and
high expression states. In this paper, we explore the basis of bimodal
protein distributions in the three scenarios outlined above. The results
are interpreted in terms of the development of genetic competence
in \emph{B. subtilis} bacterial population.

\begin{figure}
\begin{center}\includegraphics{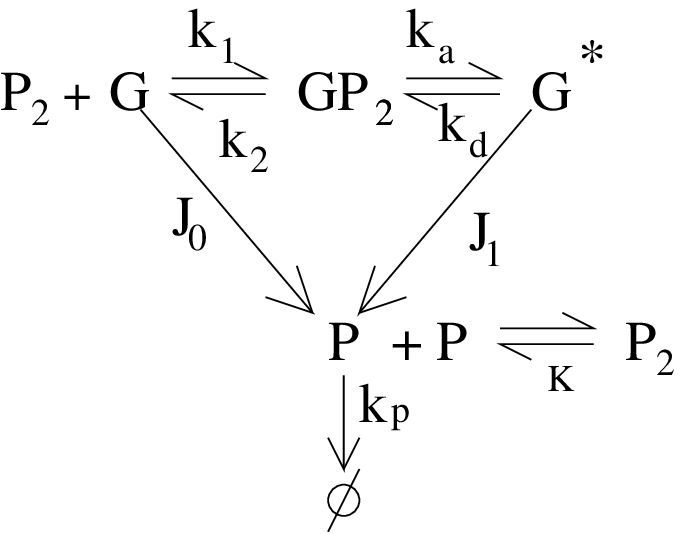}\end{center}

FIG. 1(a): The kinetic scheme describing autoregulatory gene expression.
$G$ and $G^{*}$are the inactive and active states of the gene. In
the inactive state, proteins are synthesized at a basal rate $J_{0}$.
The protein molecules form dimers $P_{2}$ which bind to the promoter
region of the gene and activate the state $G$ to $G^{*}.$

\begin{center}\includegraphics{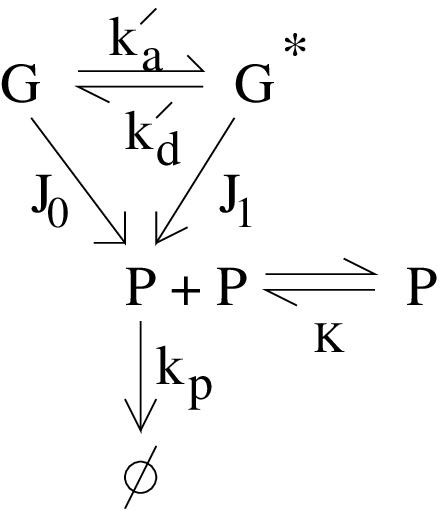}\end{center}

FIG. 1(b): The reduced kinetic scheme with effective activation and
inactivation rate constants $k_{a}^{'}(x)$ and $k_{d}^{'}$.
\end{figure}

\section{Deterministic model}

We consider a simple model of autoregulatory gene expression involving
a single gene. The proteins synthesized by the gene form dimers. The
dimer molecules bind to the promoter region of the gene and activate
gene expression, thus constituting a positive feedback loop. Apart
from autoactivation, the gene synthesizes proteins at a basal level.
The detailed kinetic scheme of the model is shown in figure 1(a).
The gene can be in two possible states $G$ (inactive) and $G^{*}$(active).
In the active state proteins are synthesized with rate constant $J_{1}.$
In the inactive state $G$, {}``leaky'' gene expression occurs at
the basal rate $J_{0}$ ($J_{1}>>J_{0}$). The basal rate may be enhanced
using appropriate inducer molecules. The synthesized proteins dimerize
with $K$ being the equilibrium dissociation constant. The protein
dimer $P_{2}$ binds to the gene in its inactive state $G$ and activates
the gene to the state $G^{*}$. The rate constants $k_{a}$ and $k_{d}$
are the activation and deactivation rate constants. The synthesized
proteins are degraded with a rate constant $k_{p}$. The kinetic scheme
in figure 1(a) can be mapped onto a simpler scheme shown in figure
1(b). The effective activation and deactivation rate constants $k_{a}^{'}(x)$
and $k_{d}^{'}$ are given by\begin{equation}
k_{a}^{'}(x)=k_{a}\frac{(x/k_{s})^{2}}{1+(x/k_{s})^{2}},\qquad\qquad k_{d}^{'}=k_{d}\label{eq:1}\end{equation}

\noindent where $x$ denotes the protein concentration and $k_{s}=\sqrt{\frac{k_{2}}{k_{1}}K}$.

In the simplified kinetic scheme of figure 1(b), the rate of change
of protein concentration is given by \begin{equation}
\frac{dx}{dt}=\frac{J_{1}k_{a}^{'}(x)}{k_{a}^{'}(x)+k_{d}}+\frac{J_{0}k_{d}}{k_{a}^{'}(x)+k_{d}}-k_{p}x\label{eq:2}\end{equation}

\noindent In the steady state, $\frac{dx}{dt}=0$ and one can identify
a parameter region in which the system is bistable, i.e., has two
stable steady states. These states correspond to low and high values
of $x$. An unstable steady state (intermediate value of $x$) separates
the two stable steady states. Figure 2 shows a plot of $x^{s}$ versus
$J_{0}$ where $x^{s}$ denotes the steady state protein concentration.
The solid branches represent stable steady states and the dotted branch,
the unstable steady states. In a range of parameter ($J_{0}$) values,
the system is bistable. The other parameters have values $k_{a}=0.0008$,
$k_{d}=0.0005$, $k_{s}=500.0$, $J_{1}=0.1$ and $k_{p}=0.0001$
in appropriate units. Bistability is, in general, accompanied by hysteresis
\cite{key-3,key-13}. Let us assume that the system is in the lower
steady state and the value of $J_{0}$ is small. As $J_{0}$ is increased
(say, with the help of inducer molecules), the system continues to
be in the low expression state. At a critical value $J_{0UC}$, a
discontinuous transition to the upper stable steady state occurs.
If $J_{0}$ is increased further, the system is monostable, i.e.,
there is only one stable steady state (the upper state). If the value
of $J_{0}$ is now reduced below $J_{0UC}$, the system remains in
the upper steady state which is a hallmark of hysteresis. At a lower
critical value of $J_{0}=J_{0LC}$ (marked by a vertical line on the
horizontal axis of figure 2), a transition from the upper to the lower
stable steady state occurs. 

Hysteresis promotes robustness as once the system is in the upper
stable steady state, small fluctuations of $J_{0}$ around $J_{0UC}$
will not give rise to a transition to the lower stable steady state.
Let us now assume that the basal gene expression is initiated with
the help of inducer molecules. These molecules may have a heterogeneous
distribution in the cell population (each individual cell contains
the autoregulatory module) which gives rise to a distribution in the
basal rates $J_{0}$. If the threshold value $J_{0UC}$ falls within
the $J_{0}$ distribution, the cell population exhibits bimodality.
Cells in which the basal rate $J_{0}$ is less (greater) than $J_{0UC}$,
are in the low (upper) stable steady state. Figures 3(a) and (b) illustrate
this for a normal distribution of basal rates with mean $=0.00445$
and variance $=0.0005$. In figure 3(b), $p(x)$ describes the steady
state distribution in the protein levels. In the steady state, $\frac{dx}{dt}=0$
in equation (\ref{eq:2}), from which the basal protein synthesis
rate $J_{0}$ can be expressed as a function of $x$ i.e., $J_{0}=f(x)$.
Let $p(J_{0})$ be the distribution in basal levels ($J_{0}/k_{p}$
is the steady state basal level). One can then write\begin{equation}
p(x)=p(j_{0})\mid_{J_{0}=f(x)}\:\mid\frac{dJ_{0}}{dx}\mid\label{eq:3}\end{equation}

\noindent This way of explaining bimodality is consistent with an
earlier proposal on the origin of binary gene expression \cite{key-14}.
We now discuss the other two mechanisms for obtaining bimodality taking
stochasticity in gene expression explicitly into account.

\begin{figure}
\begin{center}\includegraphics{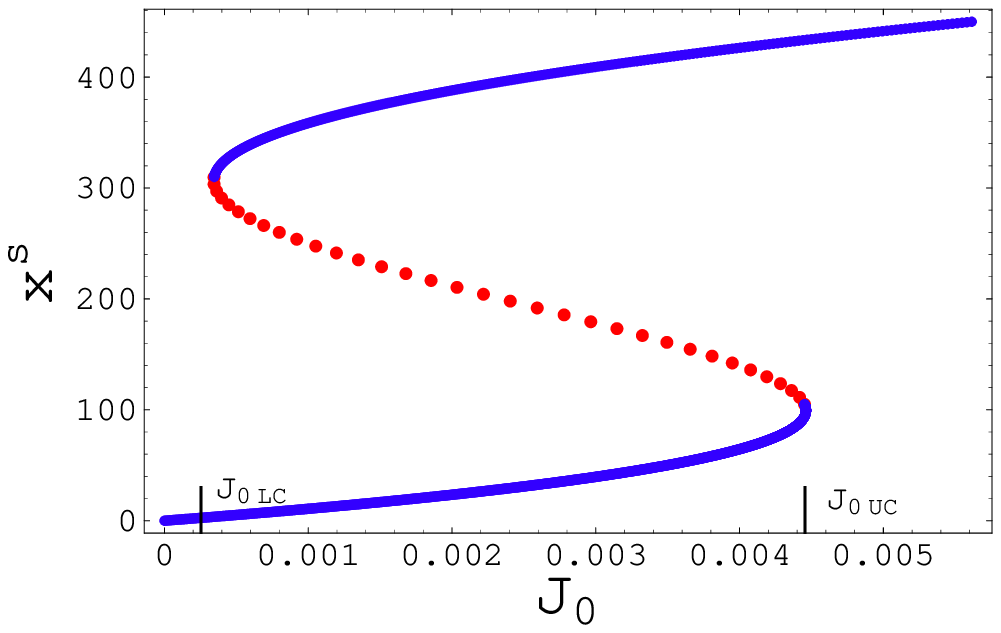}\end{center}

FIG. 2: Bistability and hysteresis, the solid (dotted) lines represent
stable (unstable) steady states; $x^{s}$ is the steady state concentration
of proteins and $J_{0}$, the basal rate of protein synthesis, serves
as the bifurcation parameter. The short vertical lines on the horizontal
axis denote the lower and upper bifurcation points, $J_{0LC}$ and
$J_{0UC}$. 
\end{figure}
\begin{figure}
\begin{center}\includegraphics{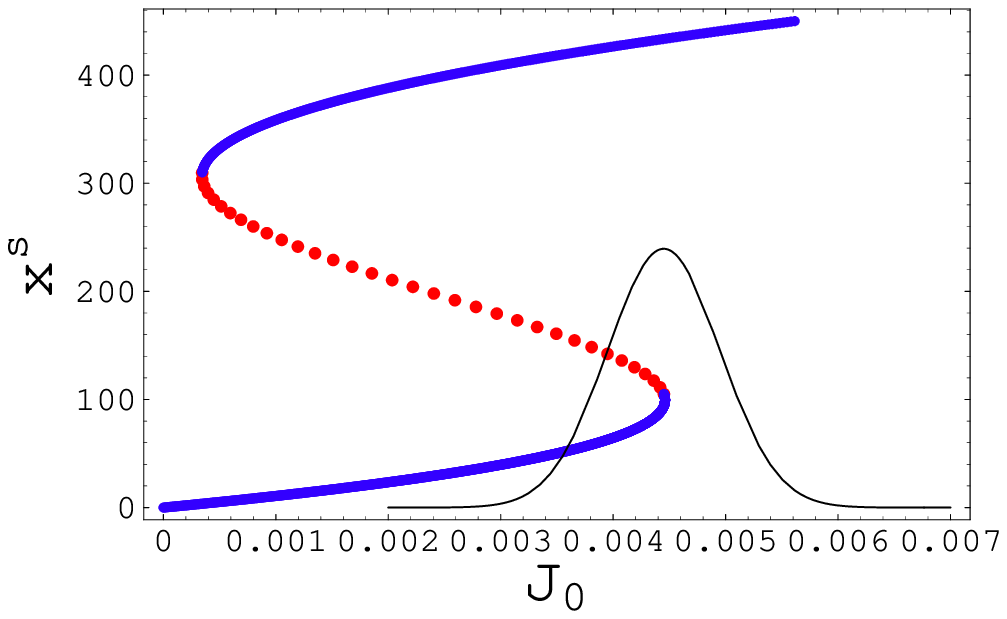}\end{center}

FIG. 3(a): Normal distribution describing heterogeneous inducer distribution
overlaps with the upper bifurcation point $J_{0UC}$.

\begin{center}\includegraphics{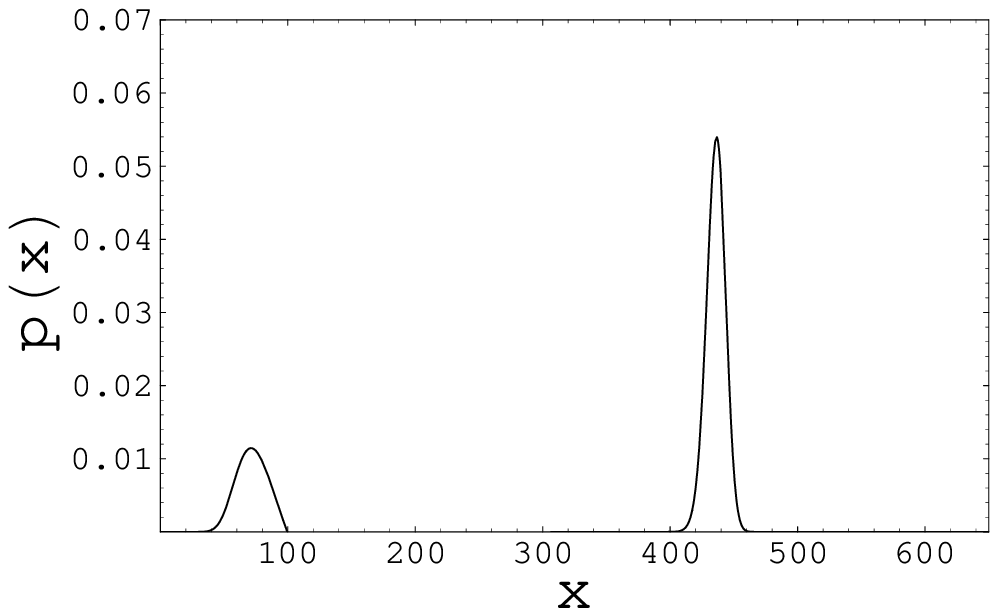}\end{center}

FIG. 3(b): The bimodal distribution in protein levels due to the heterogeneous
distribution of inducer molecules.
\end{figure}

\section{Stochastic origins of bimodality}

We consider a simple stochastic model corresponding to the kinetic
scheme in figure 1(b). In the model, the only stochasticity arises
from the random transitions of the gene between the inactive and active
states as in the minimal model of Cook et al. \cite{key-15}. Protein
synthesis from the inactive (basal expression) and active states of
the gene and protein degradation occur in a deterministic manner.
We would like to determine the distribution of protein levels in the
steady state of the cell population. Following the method outlined
in \cite{key-16}, the concentration of proteins evolves as \begin{equation}
\frac{dx}{dt}=J_{1}z+J_{0}(1-z)-k_{p}x=f(x,z)\label{eq:4}\end{equation}

\noindent where $z=1(0)$ when the gene is in the active, $G^{*}$(inactive,
$G$) state. The random variable $z$ switches values with stochastic
rate constants $k_{a}^{'}(x)\:(0\rightarrow1)$ and $k_{d}^{'}\:(1\rightarrow0)$.
Let $p_{j}(x,t)$ $(j=0,$ $1)$ be the probability density function
when $z=j.$ The total probability density function is \begin{equation}
p(x,t)=p_{0}(x,t)+p_{1}(x,t)\label{eq:5}\end{equation}

\begin{flushleft}The rate of change of probability density is given
by\begin{equation}
\frac{\partial p_{j}(x,t)}{\partial t}=-\frac{\partial}{\partial x}[f(x,j)\: p_{j}(x,t)]+\sum_{k\neq j}[W_{kj}\: p_{k}(x,t)-W_{jk}\: p_{j}(x,t)]\label{eq:6}\end{equation}
\end{flushleft}

\noindent where $W_{kj}$ is the transition rate from the state $k$
to the state $j$ and $W_{jk}$ is the same for the reverse transition.
The first term in equation (\ref{eq:6}) is the so called {}``transport''
term representing the net flow of the probability density. The second
term represents the gain/loss in the probability density due to random
transitions between the state $j$ and other accessible states. In
the present case, equation (\ref{eq:6}) gives rise to the following
two equations:\begin{equation}
\frac{\partial p_{0}(x,t)}{\partial t}=-\frac{\partial}{\partial x}\{(J_{0}-k_{p}\: x)p(x,t)]\}+k_{d}\: p_{1}(x,t)-k_{a}^{'}(x)\: p_{0}(x,t)\label{eq:7}\end{equation}

\begin{equation}
\frac{\partial p_{1}(x,t)}{\partial t}=-\frac{\partial}{\partial x}\{(J_{1}-k_{p}\: x)\: p_{1}(x,t)\}+k_{a}^{'}(x)\: p_{0}(x,t)-k_{d}\: p_{1}(x,t)\label{eq:8}\end{equation}

\noindent Using equation (\ref{eq:5}), the steady state solution
of equations (\ref{eq:7}) and (\ref{eq:8}) is given by 

\begin{flushleft}\begin{equation}
p(x)=C\:(k_{p}\: x-J_{0})^{-v}(J_{1}-k_{p}\: x)^{-1+\frac{k_{d}}{k_{p}}}(x^{2}+k_{s}^{2})^{w}Exp[u]\label{eq:9}\end{equation}
\end{flushleft}

\noindent where\begin{equation}
u=\frac{k_{a}\arctan(x/k_{s})\: J_{0}\: k_{s}}{J_{0}^{2}+k_{p}^{2}\: k_{s}^{2}},\: v=\frac{J_{0}^{2}(-k_{a}+k_{p})+k_{p}^{3}\: k_{s}^{2}}{k_{p}(J_{0}^{2}+k_{p}^{2}\: k_{s}^{2})},\: w=\frac{k_{a}\: k_{p}\: k_{s}}{2(J_{0}^{2}+k_{p}^{2}\: k_{s}^{2})}\label{eq:10}\end{equation}

\noindent and $C$ is the normalization constant.

Figures 4(a)-(d) show the plots of $p(x)$ versus $x$ as the activation
and deactivation rate constants $k_{a}=0.0008*h$ and $k_{d}=0.0005*h$
are progressively changed by varying the factor $h$. The other parameters
have values $k_{s}=500,$ $J_{0}=0.0035,$ $J_{1}=0.1$ and $k_{p}=0.0001.$
As $h$ is changed from $2$ to $100$, there is a transition from
unimodality (figure 4(a)) to bimodality (figures 4(b) and (c)) to
again unimodality (figure 4(d)). The unimodal distributions correspond
to low (figure 4(a)) and high (figure 4(d)) gene expression states.
In all the four cases, the deterministic dynamics (equation (\ref{eq:2}))
lead to bistability in the steady state (figure 2). The three steady
state solutions for $J_{0}=0.0035$ are $x_{stable1}=50.11,$ $x_{unstable}=160.60$
and $x_{stable2}=418.13$. The bimodality observed in figures 4(b)
and (c) are due to stochastic transitions between the stable steady
states brought about by the fluctuations associated with the protein
levels. The two stable steady states $x_{stable1}$ and $x_{stable2}$
are separated by the unstable steady state $x_{unstable}$. When protein
levels are below (above) $x_{unstable}$, the low (high) expression
state becomes the stable steady state. Fluctuations in the protein
levels (due to stochastic gene expression) are responsible for the
excursions from one state to the other. In the deterministic picture
(figure 2), the system is in the lower stable steady state for the
parameter value $J_{0}=0.0035$ used in obtaining the plots in figure
4. Figures 4(b)-(d) thus clearly demonstrate that noise can alter
the deterministic outcome in a significant manner. In case (d), the
fluctuations associated with the lower protein level are so strong
that excursions to the higher protein level occur with probability
one. In terms of the autoregulatory genetic module, the autostimulatory
feedback loop is fully activated when the protein level $x$ is $>x_{unstable}$
so that the high expression state is achieved in the steady state.
In the deterministic picture, the time evolution of a dynamical system
can be predicted with absolute certainty once the parameter values
and the initial state are specified. In the present case, the different
rate and binding constants constitute the parameters. The state of
the system at time $t$ is given by the amount of proteins $x(t).$
The value of $x(t)$ is obtained by solving the differential equation
(equation (2)) for a fixed set of parameter values and with a knowledge
of the initial state $x(t_{0})$ at time $t_{0}$. The time evolution
of the system is represented by a trajectory in state space (one-dimensional
in the present case). The trajectory starts from the point $x(t_{0})$and
ends at a fixed point ($\frac{dx}{dt}=0$) describing a stable steady
state. In the region of bistability, the two stable steady states
$x_{stable1}$ and $x_{stable2}$ have their individual basins of
attraction \cite{key-17,key-18}. A trajectory which starts in one
particular basin of attraction reaches the corresponding stable steady
state in the course of time. The time evolution of a system stops
once the steady state is reached. A steady state is stable (unstable)
if the system comes back to it after a weak perturbation is applied.
Small fluctuations in the protein level $x$ leave the system in the
same basin of attraction. There may, however, be excursions from one
basin to the other when the fluctuations are of sufficiently large
magnitude. The probability of transition from one basin of attraction
to the other depends amongst other factors on the value of $J_{0}$,
the basal rate of protein synthesis. The gap between $x_{unstable}$
and $x_{stable1}$ is smaller and that between $x_{stable2}$ and
$x_{unstable}$ larger as $J_{0}$ approaches $J_{0UC}$. The reverse
situation is true as $J_{0}$ approaches the lower bifurcation point.
The plots in figure 4 have been obtained for progressively higher
values of the activation rate constant $k_{a}$. The value of $J_{0}=0.0035$
is closer to the upper bifurcation point $J_{0UC}$. The protein fluctuations
are amplified for higher values of $k_{a}$. The fluctuations have
to bridge a smaller gap for transition from the basin of attraction
of $x_{stable1}$to that of $x_{stable2}$ than in the case of the
reverse transition. In the case of figure 4(a), the system remains
in the basin of attraction of $x_{stable1}$. As $k_{a}$ is made
higher, a greater fraction of the cell population attains the high
expression state. In the case of figure 4(d), almost the whole cell
population is in the high expression state. Figures 5(a)-(d) show
plots similar to those in figure 4 for a lower value of $J_{0}=0.0030$.
The gap between $x_{unstable}$ and $x_{stable1}$ is now larger and
that between $x_{stable2}$ and $x_{unstable}$ smaller (see figure
2). The balance in this case tilts in the favour of the lower stable
steady state. 

\begin{figure}
\includegraphics[%
  scale=0.75]{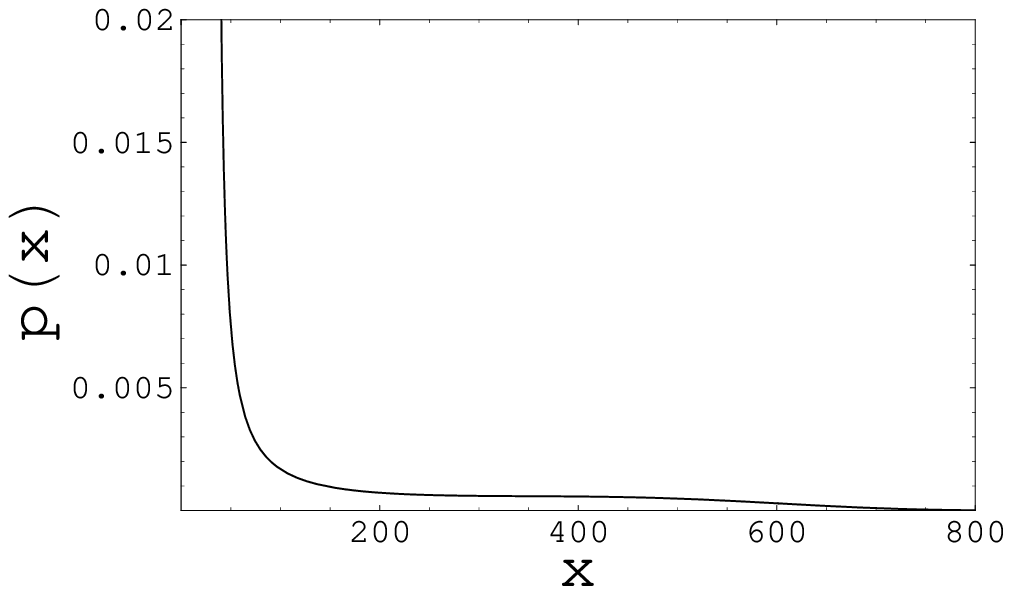}(a)\includegraphics[%
  scale=0.75]{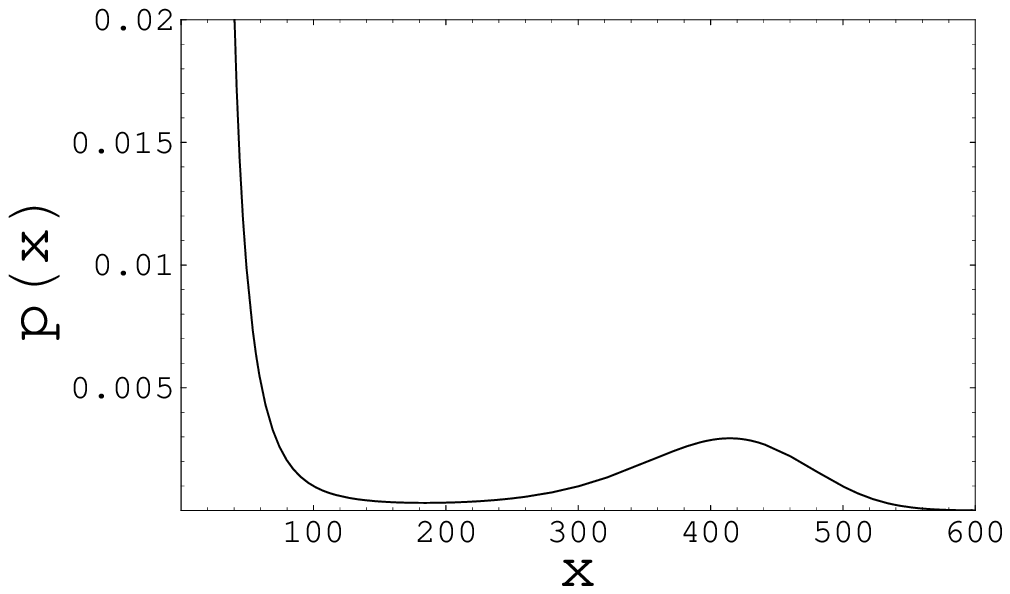}(b)

\includegraphics[%
  scale=0.75]{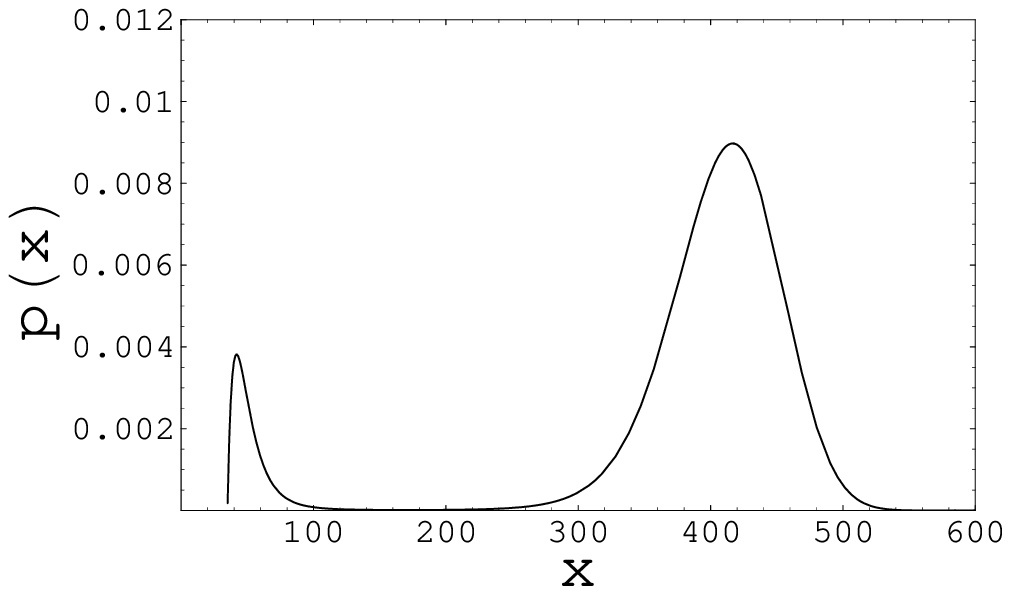}(c)\includegraphics[%
  scale=0.75]{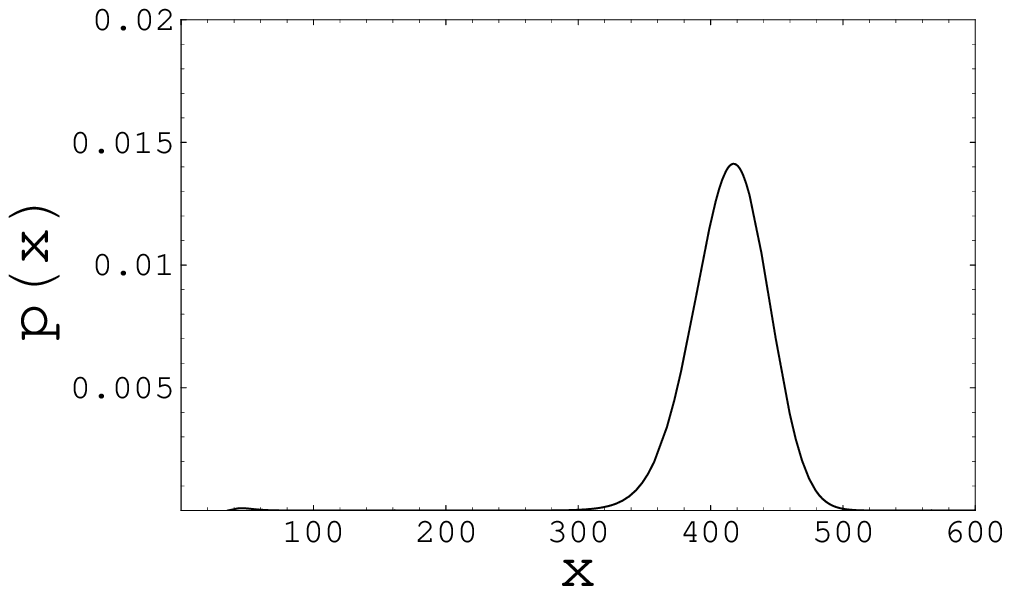}(d)

FIG. 4: The steady state distribution, $p(x)$versus $x$, in protein
levels with the activation and inactivation rate constants given by
$k_{a}=0.0008*h$ and $k_{d}=0.0005*h$. The plots are obtained for
different values of $h$, (a) $h=2$, (b) $h=20$, (c) $h=50$ and
(d) $h=100$. The basal rate of protein synthesis is $J_{0}=0.0035$.
The three steady states in the deterministic case are $x_{stable1}=50.11$,
$x_{unstable}=160.6$ and $x_{stable2}=418.13$.
\end{figure}
\begin{figure}
\includegraphics[%
  scale=0.75]{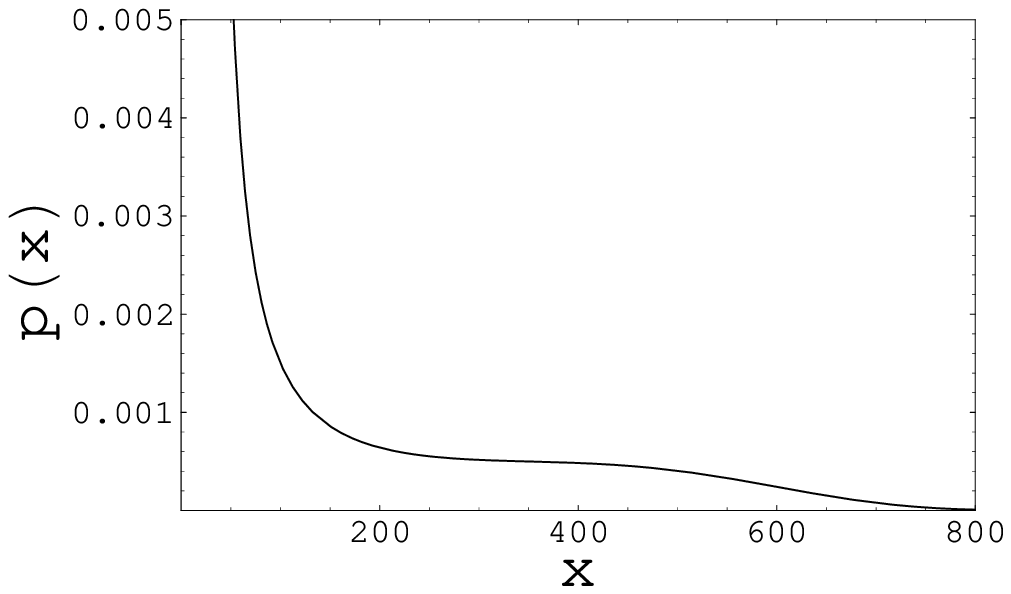}(a)\includegraphics[%
  scale=0.75]{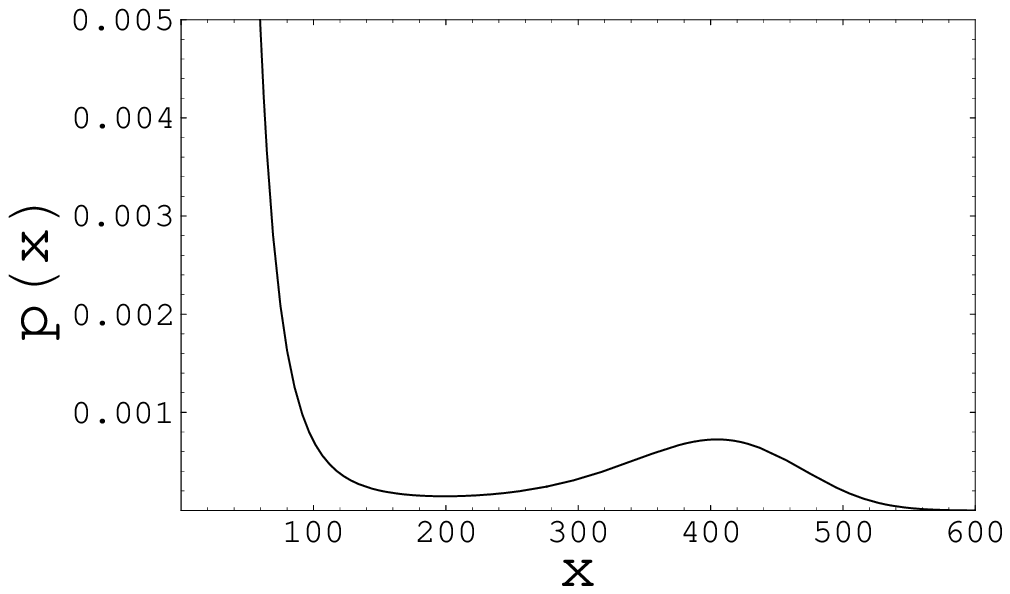}(b)

\includegraphics[%
  scale=0.75]{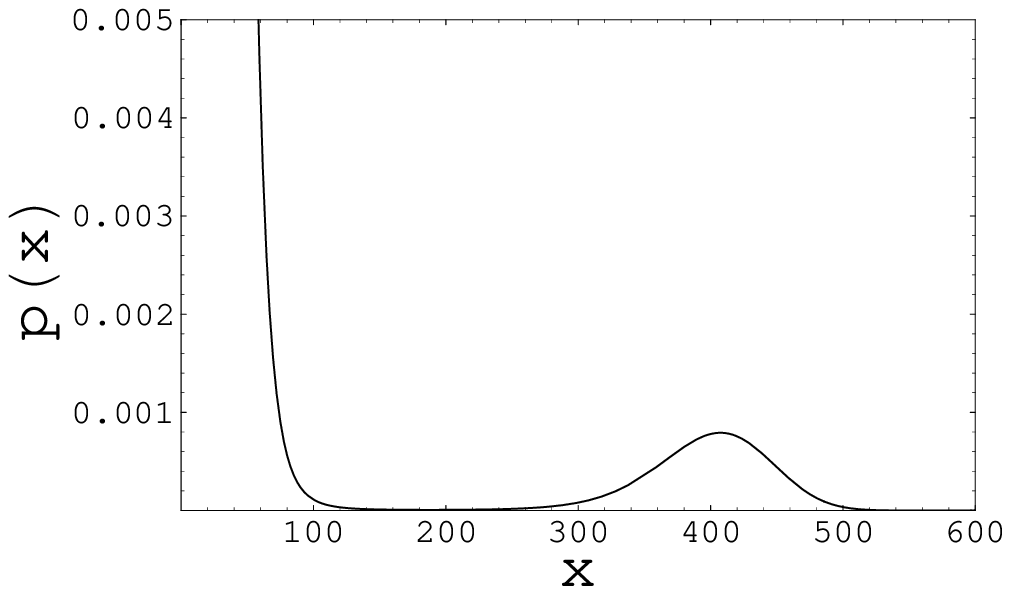}(c)\includegraphics[%
  scale=0.75]{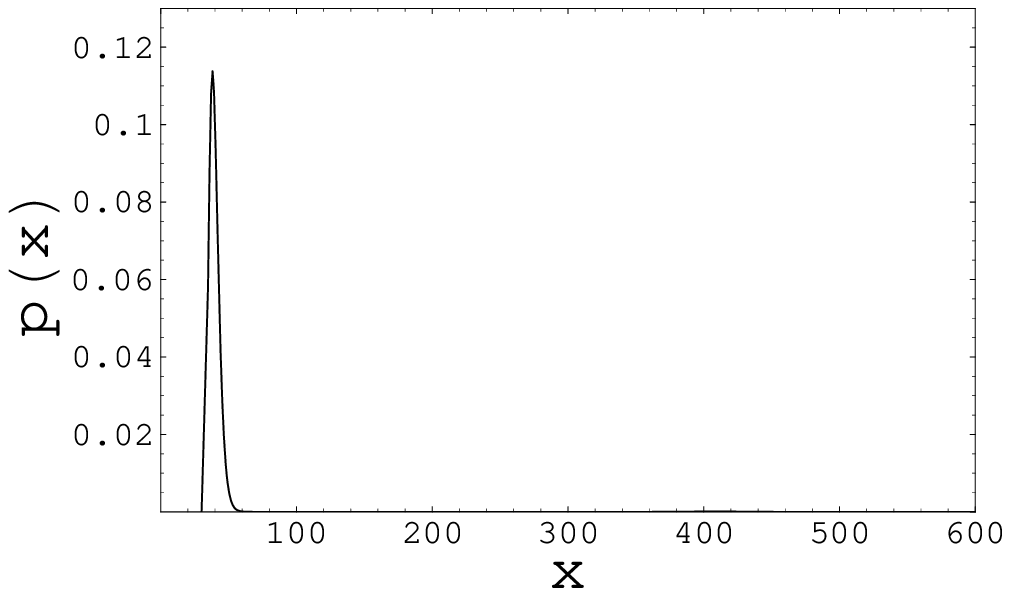}(d)

FIG. 5: The steady state distribution, $p(x)$ versus $x$, in protein
levels. The parameter $J_{0}$ has the value $J_{0}=0.0030$ with
$k_{a}=0.0008*h$ and $k_{d}=0.0005*h$ as in the case of figure 4.
The values of $h$ are (a) $h=2$, (b) $h=20$, (c) $h=50$ and (d)
$h=250$. The three steady states in the deterministic case are $x_{stable1}=39.55$,
$x_{unstable}=178.22$ and $x_{stable2}=409.14$.
\end{figure}

We now consider the third case in which a bimodal protein distribution
has a purely stochastic origin. The system is monostable in the deterministic
description. An earlier study by Kepler and Elston \cite{key-19}
provides examples of such cases. Some other studies have explored
the basis of stochastic binary gene expression in different settings
(without positive feedback) \cite{key-8,key-12,key-20,key-21}. An
example in the case of autoregulated gene expression is shown in figure
6 for the parameter values $k_{a}=0.0012$, $k_{d}=0.0004$, $k_{s}=500.0,$
$J_{0}=0.01,$ $J_{1}=0.1$ and $k_{p}=0.0001.$ In the deterministic
description, there is only one stable steady state, $x^{s}=581.3$.
The protein distribution is obtained from the analytic expression
given in equation (9). The stochastic model considered in this section
is analytically tractable because of certain simple assumptions. The
only stochasticity considered in the model is that associated with
random gene activation and deactivation. The autocatalytic feedback
is incorporated in an effective rate constant $k_{a}^{'}(x)$. The
gene expression is considered as a one-step process, i.e., the intermediate
stage of $mRNA$ synthesis is not explicitly taken into account. We
now describe the results of a detailed simulation based on the Gillespie
algorithm \cite{key-22} which takes the two-step nature of gene expression
into account and treats the distinct biochemical events to be stochastic
in nature. The different biochemical reactions are listed in equations
(\ref{eq:11})-(\ref{eq:21}):\begin{equation}
G+P2\rightarrow GP2\label{eq:11}\end{equation}
\begin{equation}
GP2\rightarrow G+P2\label{eq:12}\end{equation}
\begin{equation}
GP2\rightarrow G*\label{eq:13}\end{equation}
\begin{equation}
G*\rightarrow GP2\label{eq:14}\end{equation}
\begin{equation}
G*\rightarrow m\label{eq:15}\end{equation}
\begin{equation}
G\rightarrow m\label{eq:16}\end{equation}
\begin{equation}
m\rightarrow deg.\label{eq:17}\end{equation}
\begin{equation}
m\rightarrow P\label{eq:18}\end{equation}
\begin{equation}
P\rightarrow deg.\label{eq:19}\end{equation}
\begin{equation}
P+P\rightarrow P2\label{eq:20}\end{equation}
\begin{equation}
P2\rightarrow P+P\label{eq:21}\end{equation}

\noindent In the above equations, the mRNA and protein are represented
by $m$ and $P$ respectively, $P2$ is a protein dimer and $GP2$
denotes the intermediate state of a protein dimer bound to the gene
in its inactive state $G$. Equations (\ref{eq:17}) and (\ref{eq:19})
describe the degradation of the mRNA and protein molecules. In the
simulation, the stochastic rate constants associated with the equations
(\ref{eq:11})-(\ref{eq:21}) are $c(1)=0.003$, $c(2)=0.16$, $c(3)=0.004*h$,
$c(4)=0.0006*h$, $c(5)=0.1$, $c(6)=0.0015$, $c(7)=0.0001$, $c(8)=0.000001$,
$c(9)=0.008$, $c(10)=0.01$ and $c(11)=0.01$ in appropriate units.
The simulation is carried out for three different values of $h=1$,
$10$ and $200$ respectively. The results are shown in figures 7(a)-(c).
The plots on the left show the time trajectories, $x(t)$ versus $t$,
where $x(t)$ is the amount of proteins at time $t$. The plots on
the right show the distributions $p(x)$ versus $x$ on repeating
the simulation 3000 times. The quantity $p(x)dx$ provides a measure
of the fraction of cells in a population with protein levels between
$x$ and $x+dx$. For the parameter values used in the simulation,
only a small fraction of cells is in the high expression state. Figures
7(a) and 7(b) show that the stochastic nature of the biochemical events
involved in gene expression is responsible for a bimodal protein distribution. 

\begin{figure}
\begin{center}\includegraphics{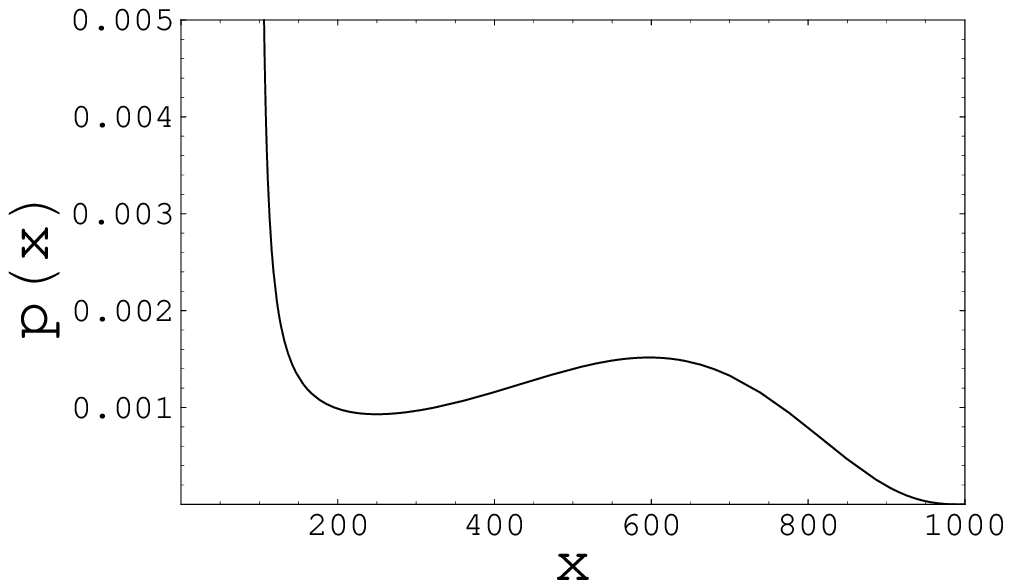}\end{center}

FIG. 6: The steady state distribution, $p(x)$ versus $x$, in protein
levels for $k_{a}=0.0008$, $k_{d}=0.0004$ and $J_{0}=0.001$. The
bimodal distribution has a purely stochastic origin. In the deterministic
case, there is only one stable steady state $x^{s}=581.3$.
\end{figure}
\begin{figure}
\includegraphics[%
  scale=0.6]{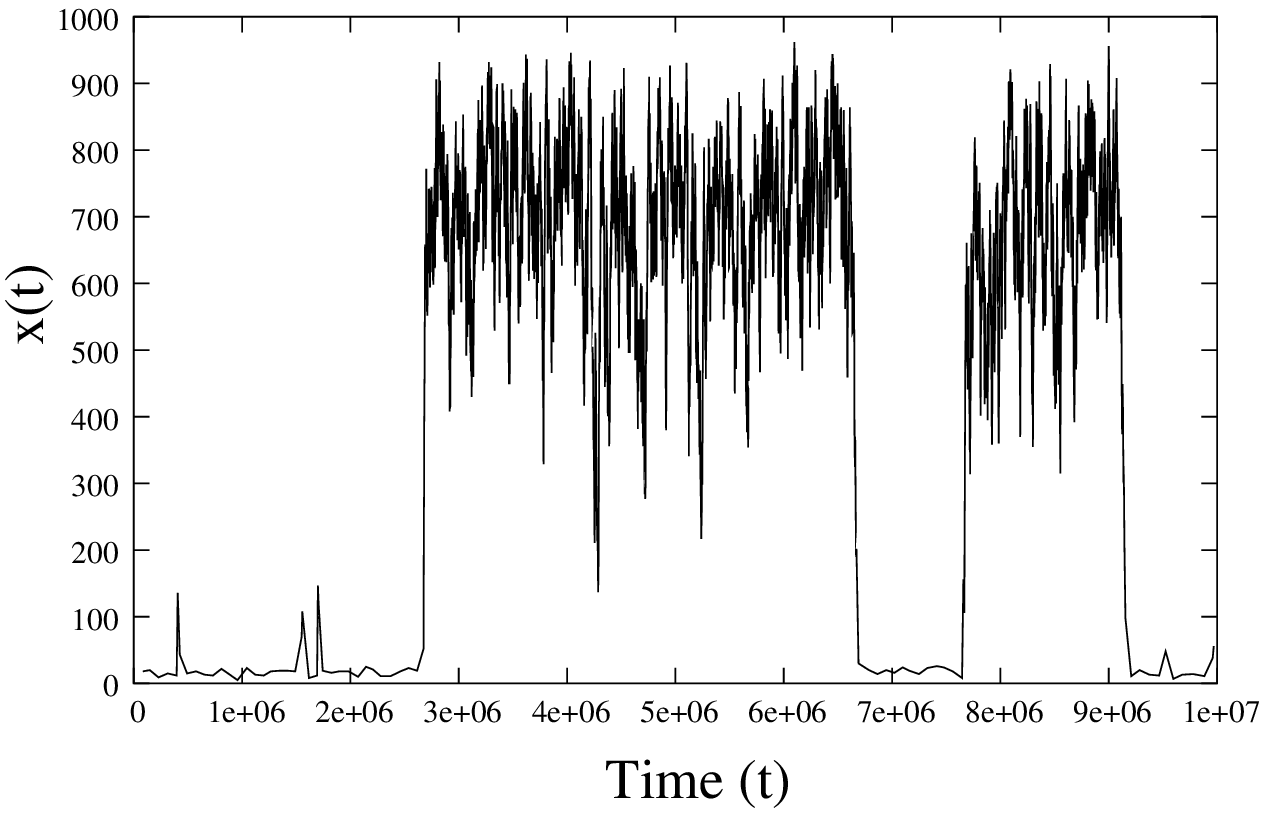}1(a)\includegraphics[%
  scale=0.6]{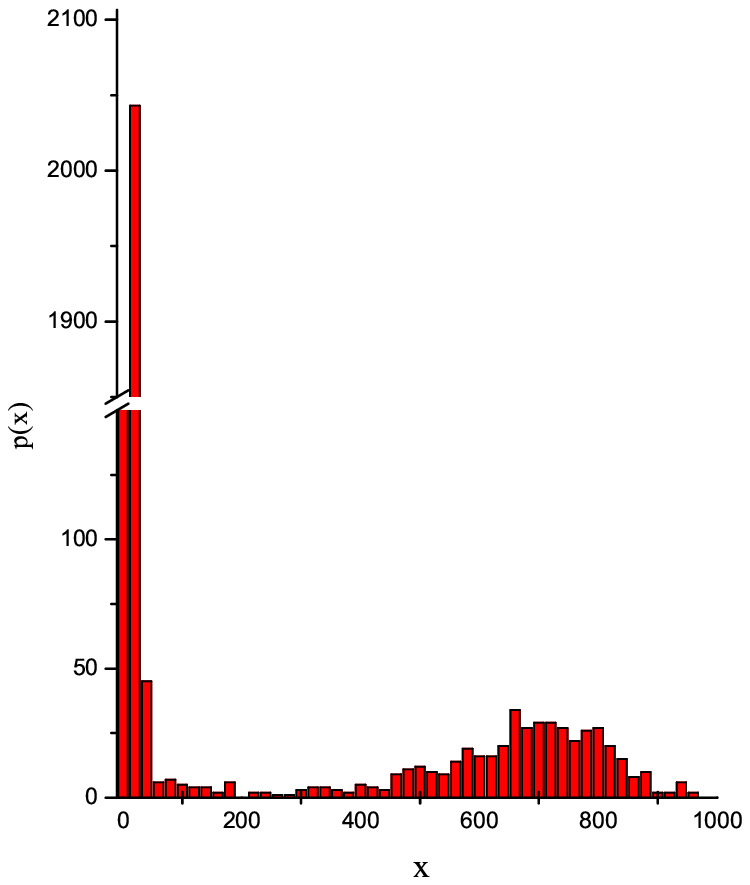}2(a)

\includegraphics[%
  scale=0.6]{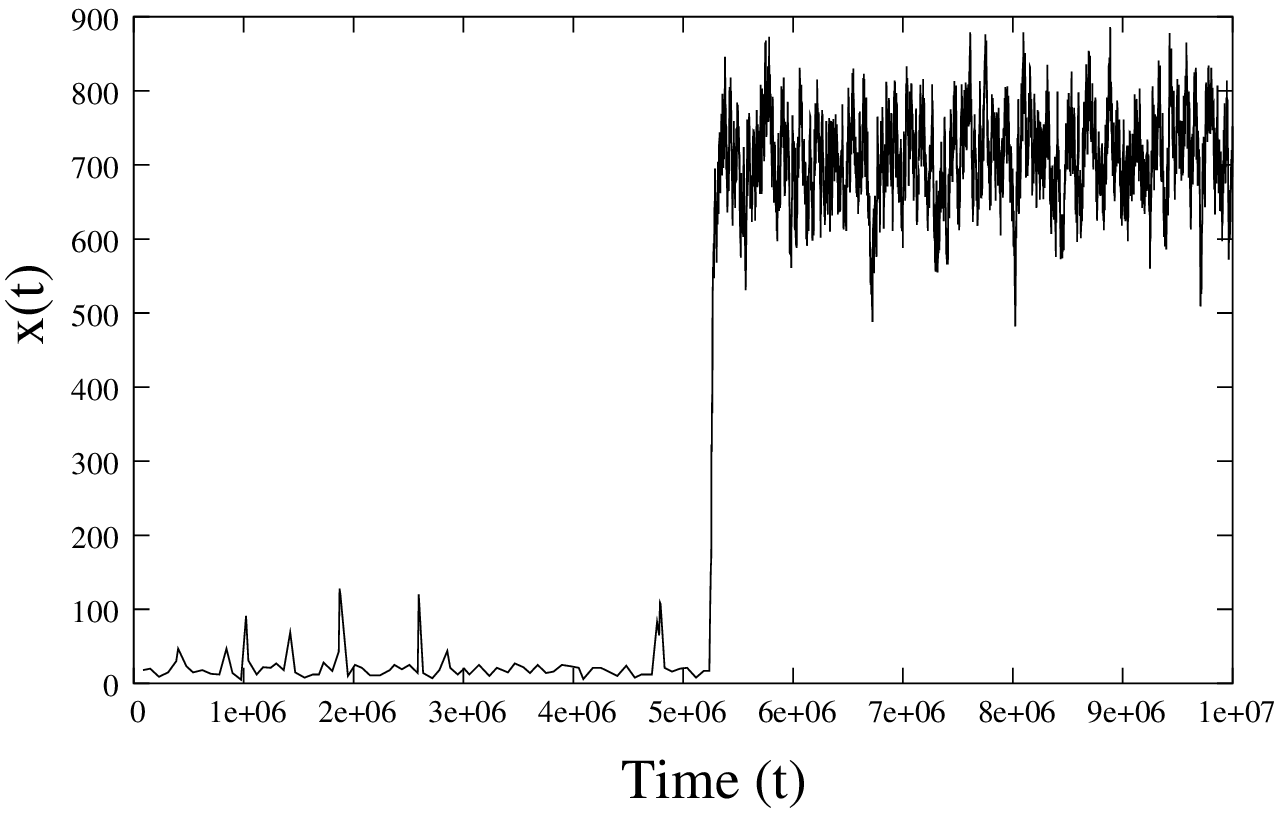}1(b)\includegraphics[%
  scale=0.6]{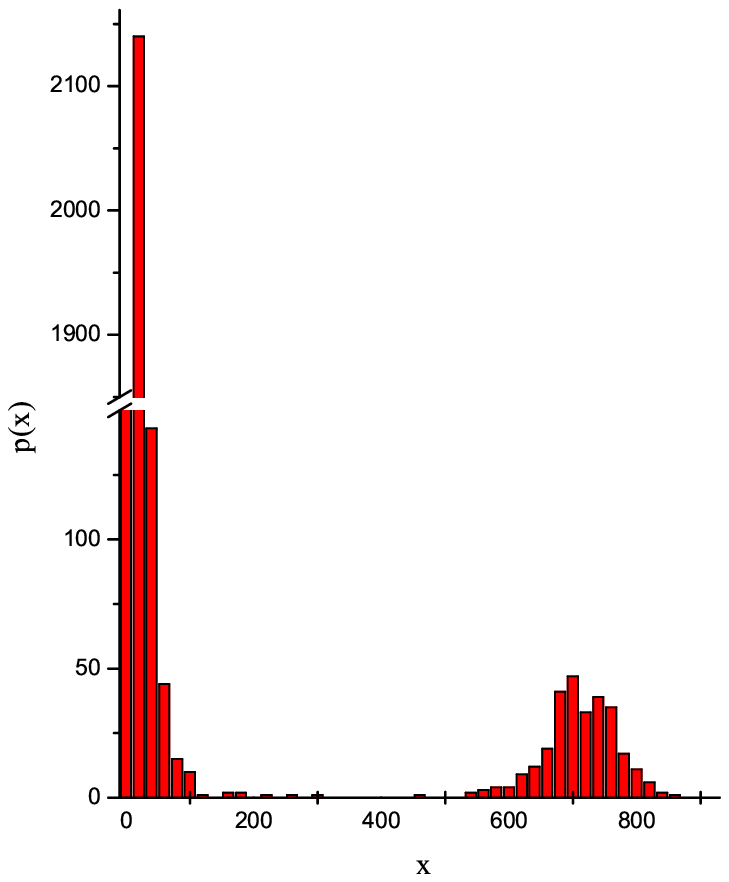}2(b)

\includegraphics[%
  scale=0.6]{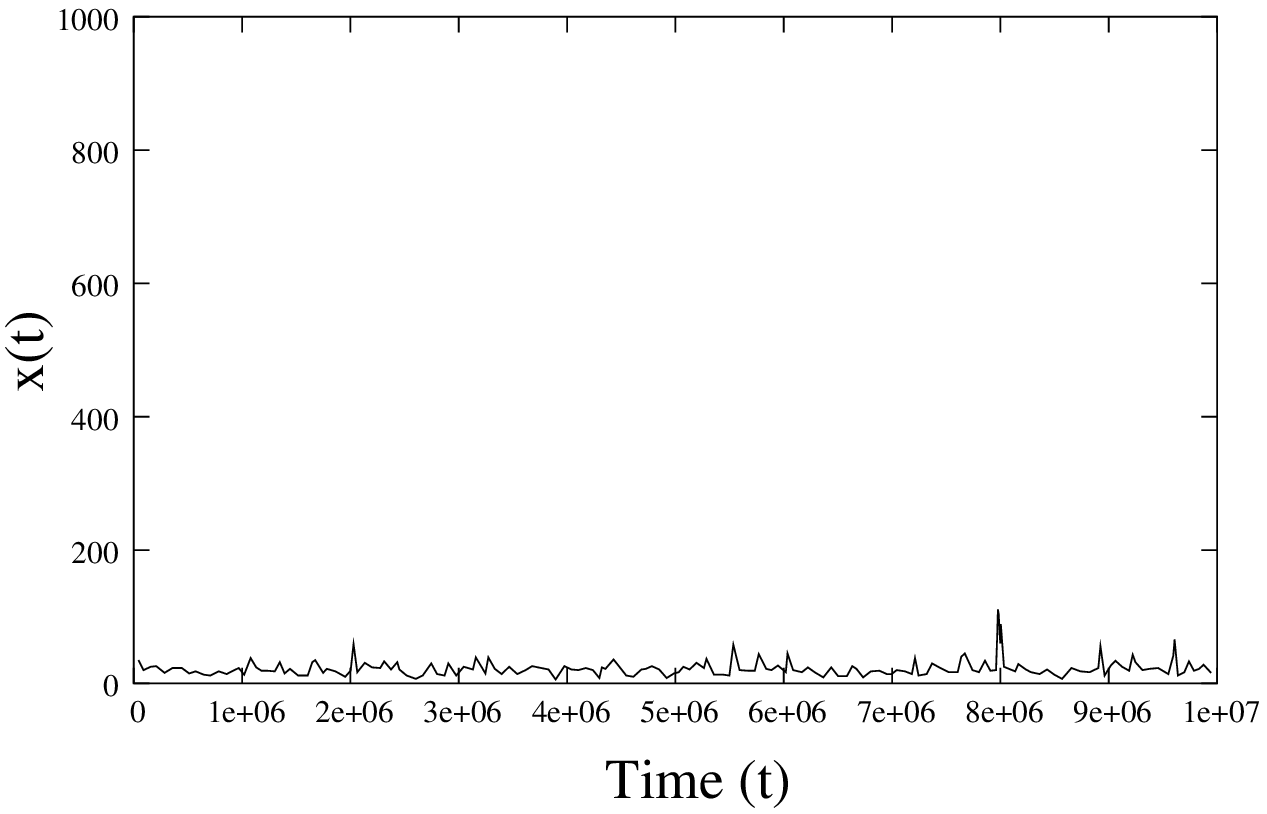}1(c)\includegraphics[%
  scale=0.6]{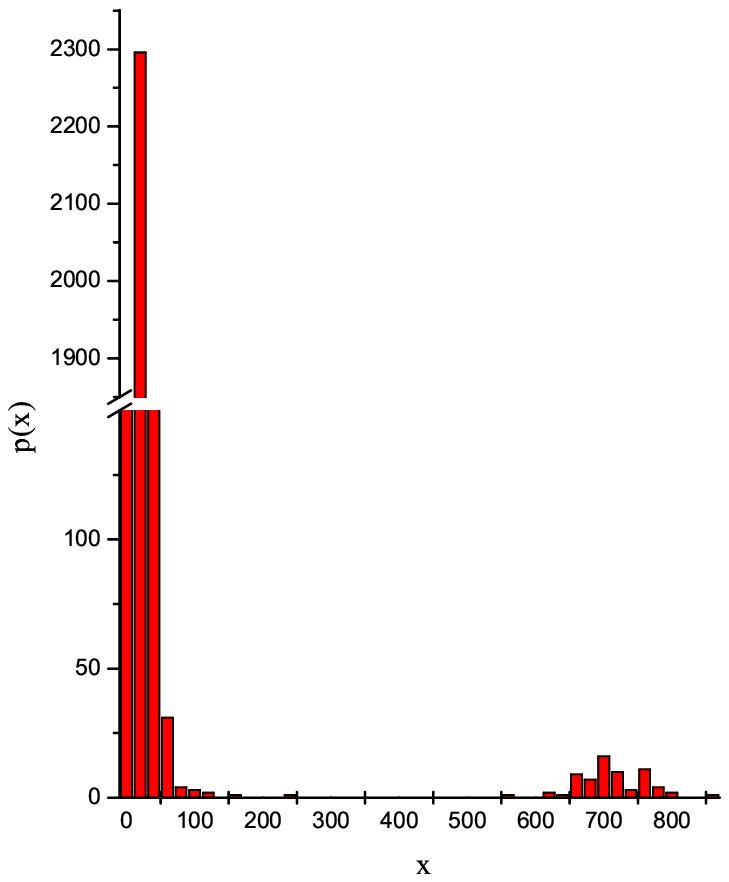}2(c)

FIG. 7: Results of simulation based on the Gillespie algorithm. The
stochastic rate constants for gene activation and inactivation are
$c(3)=0.004*h$ and $c(4)=0.0006*h$ The factor $h$ has values (a)
$h=1$, (b) $h=10$ and (c) $h=200$. The plots in the first column
show the variation of protein amount as a function of time. The plots
in the second column show the corresponding steady state distributions,
$p(x)$ versus $x$.
\end{figure}

\section{Discussions}

In this paper, we study how positive feedback combined with stochasticity
gives rise to binary gene expression, i.e., a bimodal distribution
in the protein levels in a population of cells. There are some earlier
studies \cite{key-2,key-19} on the same issue but the modeling detail
and context are different. The motivation for the present study comes
from the experimental observation that a single comK gene, which autoregulates
its expresssion via a positive feedback loop, is by itself sufficient
to generate heterogeneity in a population of \emph{B. subtilis} \cite{key-5,key-6,key-7}\emph{.}
A fraction of the cell population develops competence due to the high
expression state of comK. This is so when the ComK protein level exceeds
a threshold value thus triggering the full activation of the autostimulatory
loop. The comK autoregulatory genetic module is at the core of a complex
network of molecular interactions which regulate comK transcription
and the stability of the ComK proteins. In this case, only a small
fraction of cells, about ten percent, develops competence. When experiments
are carried out on the isolated genetic module, the fraction of cell
population in the high comK expression state can be quite large. It
has further been suggested that stochasticity in gene expression is
responsible for throwing the autocatalytic switch \cite{key-5,key-6,key-7}.
In our study, we focus only on the core module, namely, the comK autoregulatory
genetic module of the network regulating competence development. The
module represents a single gene (comK), the protein product of which
autoactivates its own synthesis through dimerization and subsequent
binding at the appropriate region of the DNA. Our model incorporates
these minimalist features of the autoregulatory comK module. We have
explored the basis of binary gene expression in the framework of three
different possibilities. In the first two cases, the underlying dynamics
lead to bistability in a deterministic description. A heterogeneous
distribution in inducer molecules may give rise to a bimodal distribution
in the protein levels. In the second and third cases, we take stochasticity
in gene expression into account and derive an analytic expression
(equation (\ref{eq:9})) for the steady state distribution of protein
levels. The analytical tractability of the stochastic model arises
from two assumptions. Firstly, the two major steps of gene expression,
namely, transcription (synthesis of mRNAs) and translation (synthesis
of proteins) are combined into a single step leading to protein production.
Secondly, the only source of stochasticity in the model lies in the
random activation and deactivation of the target gene expression.
The first assumption provides the basis for several studies of stochastic
gene expression \cite{key-15,key-19,key-21,key-23}. The second assumption
is strictly valid when the dominant source of noise is associated
with the random activation and deactivation of gene expression. This
is so in the case of slow promoter kinetics. As discussed in detail
in \cite{key-12}, slow transitions between the promoter states result
in transcriptional bursts of mRNA synthesis and increased heterogeneity
within a cell population including bimodal protein distributions.
Experimental evidence of transcriptional bursting has been obtained
for both prokaryotes and eukaryotes \cite{key-24,key-25}. A recent
experiment on stochastic mRNA synthesis in mammalian cells \cite{key-26}
shows that the mRNA levels display large cell-to-cell variations due
to random, infrequent activation of gene expression. The statistics
of the variations are adequately described by a model in which the
only source of stochasticity lies in the random activation and deactivation
of the gene. There could be a number of factors which lead to slow
transitions between the promoter states. Chromatin remodeling has
been conjectured to cause transcriptional bursts in eukaryotic systems
\cite{key-12}. In both prokaryotes and eukaryotes, pulsatile gene
expression may also result from regulatory molecules binding at and
unbinding from the DNA sites, DNA undergoing conformational changes
so that the RNA polymerase has only brief access to the promoter region
etc. In the case of \emph{E. coli}, there is experimental evidences
that long periods of inactivity are interspersed by shorter periods
when the gene is in the active state \cite{key-24}. An earlier result
of Ozbudak et al. on \emph{B. subtilis} \cite{key-8} has been reanalyzed
to show that the data are not inconsistent with the possibility of
transcriptional bursts (random gene activation-inactivation) \cite{key-24}.
Our simple stochastic model, based on random transitions between inactive
and active gene states, is thus consistent with experimental reality.
In the case of bistability, stochasticity triggers transitions between
the two stable steady states which is responsible for bimodal protein
distributions. In the case when the system is monostable in a deterministic
description, binary gene expression can still occur due to a combination
of positive feedback and stochastic transitions between the inactive
and active states of the gene. Our analytical results are supported
by the simulation results in the case of a more detailed stochastic
model in which transcription and translation are treated as separate
processes and stochasticity associated with all the biochemical steps
(equations (\ref{eq:11})-(\ref{eq:21})) are taken into account.
In both the cases, the results are valid over a wide range of parameter
values. We now briefly discuss the experimental possibility for distinguishing
between the three mechanisms discussed in the paper. As shown in figure
2, bistability implies hysteresis. A properly designed experiment
can detect hysteresis in the response ($x^{s}$ in figure 2) as the
variable along the $x-axis$ ($J_{0}$, the basal rate of protein
synthesis in figure 2) is changed. Discontinuous jumps in response
at the bifurcation points and a non-reversible response are the hallmarks
of hysteresis. The value of $J_{0}$ may be changed using appropriate
inducer molecules. One can use a cell sorter and separate a subpopulation
from a bimodal cell population. The subpopulation develops bimodality
in the course of time if there are stochastic transitions between
the low and high expression levels. 

S$\ddot{u}$el et al. \cite{key-27} have investigated competence
development on the basis of a model describing an excitable stochastic
system. The key ingredients of the model are: the comK autoregulatory
loop, the inhibition of ComK degradation by ComS proteins and repression
of the comS gene by ComK. Theoretical analysis of the model dynamics
is combined with experiments to gain insight on the entry into and
exit from the competence state. This state corresponds to an unstable
fixed point of the model dynamics. The system has only one stable
steady state in which the ComK level is low. Fluctuations in the levels
of ComK/ComS excite the system into the competence state with eventual
return to the noncompetence state. In the excitable system, repeated
stochastic triggering of the competence state is thus possible. Some
of the premises of the model like the {}``indirect'' repression
of the comS gene by ComK need experimental confirmation under wild-type
expression conditions \cite{key-28}. The study nonetheless is an
elegant example of how model studies combined with experiments can
provide a new perspective on noise-induced phenomena in biological
systems. Our model has the comK autoregulatory loop as the sole ingredient
and focuses on the specific experiment by Smits et al. \cite{key-1}
on the single autoregulatory module. Genetic competence in \emph{B.
subtilis} provides a concrete example of a natural system in which
a single gene, regulating its expression via an autoregulatory positive
feedback loop, is by itself sufficient to establish two types of stable
states in the cell population. Recently, two groups have independently
discovered a similar phenomenon in the human fungal pathogen \emph{Candida
albicans} \cite{key-29,key-30}. In both the cases, the autoregulatory
modules are parts of complex genetic circuitry. The single gene modules
almost exclusively control the cellular switch operating between two
stable states. The resulting heterogeneity is epigenetic in nature.
\emph{B. subtilis} and \emph{Candida albicans} thus illustrate the
essentiality and sufficiency of network modules in explaining particular
types of biological function. The role of the other components of
the associated regulatory networks lies in modulating the functional
response. In \emph{B. subtilis}, several genes regulate the expression
of the comK gene the protein product of which regulates the expression
of several other genes. The products of the regulatory genes modulate
the threshold for the triggering of the autocatalytic switch and influence
the stability of the ComK proteins. The additional circuitry probably
includes features which further stabilize the steady states. In \emph{Candida
albicans,} the \emph{}WOR1 gene \emph{}acts as the master regulator.
The gene autoregulates its own expression via a positive feedback
loop. The switch now operates between the cellular states: white and
opaque. The two types of cells, white and opaque, differ in their
morphologies, the genes they express, the host tissues in which they
are resident and also in their mating characteristics. In the white
cells, WOR1 is expressed at low levels whereas the levels are high
in the opaque state. As in the case of \emph{B. subtilis,} stochasticity
appears to drive the transitions between the two types of cell. The
results derived in this paper, specially those pertaining to the combined
effects of bistability and stochasticity, should be of relevance in
explaining the white-opaque switching in \emph{Candida albicans. }

\begin{center}\textbf{Acknowledgement}\end{center}

\noindent I. B. thanks W. K. Smits for some helpful discussions. R.K.
is supported by the Council of Scientific and Industrial Research,
India under Sanction No. 9/15 (239)/2002-EMR-1.

\end{document}